\def \kev{\rm{kev}}

\def \cm{~\rm{cm}}

\def \K{~\rm{K}}

\def \erg{~\rm{erg}}

\def \yr{~\rm{yr}}

\def \kpc{~\rm{kpc}}
\documentclass[12pt,preprint]{aastex}
\usepackage{natbib}
\usepackage{epsfig}

\begin{document}

\title{Cooling by Heat Conduction Inside Magnetic Flux Loops
and the Moderate Cluster Cooling Flow Model}

\author{Noam Soker\altaffilmark{1}
\altaffiltext{1}{Department of Physics, Technion$-$Israel institute
of Technology, Haifa 32000 Israel, and Department of Physics, Oranim;
soker@physics.technion.ac.il.}}

\begin{abstract}
I study non-radiative cooling of X-ray emitting gas via
heat conduction along magnetic field lines inside magnetic
flux loops in cooling flow clusters of galaxies.
I find that such heat conduction can reduce the fraction
of energy radiated in the X-ray band by a factor of $\sim 1.5-2$.
This non-radiative cooling joins two other proposed non-radiative
cooling processes, which can be more efficient.
These are mixing of cold and hot gas, and heat conduction
initiated by magnetic fields reconnection
between hot and cold gas.
These processes when incorporated into the moderate cooling
flow model lead to a general cooling flow model with
the following ingredients.
(1) Cooling flow does occur, but with a mass cooling rate
$\sim 10$ times lower than in old ($\sim 10$ years ago) versions of
the cooling flow model.
Namely, heating occurs such that the effective age of the
cooling flow is much below the cluster age, but the heating
can't prevent cooling altogether.
(2) The cooling flow region is in a non-steady state evolution.
(3) Non-radiative cooling of X-ray emitting gas can bring
the model to a much better agreement with observations.
(4) The general behavior of the cooling flow gas, and in
particular the role played by magnetic fields,
make the intracluster medium in cooling flow clusters similar
in some aspects to the active solar corona.
\end{abstract}


\section{Introduction}\label{sec:intro}

Recent {\it Chandra} and {\it XMM-Newton} observations of 
cooling flow (CF) clusters of galaxies have lead to a 
shower of papers examining reheating 
processes of the intracluster medium (ICM), which together 
with pre-{\it Chandra} and XMM-Newton reheating papers make a 
long list of papers with different ideas 
(e.g., Binney \& Tabor 1995; Tucker \& David 1997;
Ciotti \& Ostriker 2001; David et al.\ 2001;
Quilis, Bower, \& Balogh 2001; Br\"uggen, M. \& Kaiser 2001;
Ruszkowski \& Begelman 2002; Nulsen et al. 2002; Fabian 2003;
Fabian et al.\ 2003; Binney 2003; Mathews et al.\ 2003;
 for more references see Peterson et al.\ 2003b).
These recent observations, although in contradiction (Fabian 2003) 
with the cluster-CF models of 10 years ago (Fabian 1994),  are
compatible with models having low mass cooling rates (see
Binney 2003, for why the new results are not a surprise),
such as in the moderate CF model (Soker et al.\ 2001).
The moderate CF model, which was proposed before the
new results from the two X-ray telescopes, is different from many
early proposed processes which aim is to prevent the CF
in cluster of galaxies altogether.
The main ingredient of the moderate CF model is that the effective
age, i.e., the time period since the last major disturbance of 
the ICM inside the cooling radius $r_c \sim 100 \kpc$, is much
shorter than the cluster age (e.g., Binney \& Tabor 1995;
see Binney 2003; Soker et al.\ 2001).
The cooling radius is defined as the place where the radiative 
cooling time equals the cluster age. 
Originally the heating in the moderate CF model was proposed to
be intermittent (Soker et al.\ 2001),
but the basic idea can hold for a steady heating, or heating
in short intervals (Binney 2003).

Kahn et al. (2003) and Peterson et al. (2003a), among others, find 
that there is less gas than that predicted by old versions of
the steady-state isobaric CF model at all temperatures below 
the ambient gas temperature. 
Also, the discrepancy increases with decreasing temperature.  
There is not just a deficit in gas below $\sim 1/3$ 
of the ambient cluster temperature, there is a deficit of gas at all 
temperatures below the ambient temperature (Peterson et al.\ 2003b).
Soker \& David (2003) show that these XMM-Newton results are consistent 
with the expectations of the moderate CF model.
Soker \& David (2003) claim that the X-ray emitting gas in CF clusters
can only be in steady-state if there exists a steady heating mechanism
that scales as $H(T) \propto T^{\alpha}$ where $\alpha=1-2$; this is 
unlikely.
They show then that in the moderate CF model most of the gas within CF 
resides in the hottest gas, which is prevented from cooling 
continuously and attaining a steady-state configuration by being 
reheated.
This results in a mass cooling rate that decreases with decreasing
temperature, with a much lower mass cooling rate at the lowest
temperatures, as discovered by Peterson et al. (2003a).
Although there is no indication yet for a violent heating, I
note the following. 
An intermittent reheating can work for an AGN kinetic energy of 
$\sim 10^{47}~{\rm erg}~{\rm s}^{-1}$, and its strong activity
should last $\sim 1-2 \times 10^7$ yr and occur every $\sim 10^9$ yr
(Soker et al.\ 2001).
A violent heating event by AGN jets was proposed by Kawano, Ohto, 
\& Fukazawa (2003) to have occurred $>10^9$ yr ago in the CF clusters
2A 0335 + 096 and A 2199. 
Mazzotta, Edge, \& Markevitch (2003) claim that the structure of the 
cluster 2A 0335+096 can be accounted for if strong AGN activity periods 
are followed by relative quiescence periods.
O'Sullivan \& Vrtilek (2003) propose that a heating event occured
$\sim 2 \times 10^8 \yr$ ago in the poor cluster MKW 4, which  
consequently is presently not in a steady state cooling.
Such a scenario is in accord with the explanation of Soker \& David (2003)
to the observations of Peterson et al.\ (2003a) and Kahn et al.\ (2003). 
Only $\sim 1-2 \%$ of all CF clusters should be found during that stage.
In addition, Heinz \& En{\ss}lin (2003) argue for bias against detecting
strong shocks. 
Therefore, from the well studied CF clusters, only one, or at most
two, are expected to be in the stage of violent heating. 
In Cygnus A there is a strong radio source which heats the
ICM (Smith et al. 2002).
Wilson, Young, \& Smith (2003) estimate the mechanical power of
the jets in Cygnus A to be
$L_{\rm jet} \simeq 6 \times 10^{46} ~{\rm erg}~{\rm s}^{-1}$.
This is $\sim 100$ times larger than the radio emission,
and it is in the range required by the moderate CF model.
Therefore, present observations do not rule-out a violent intermittent
reheating process acting for a short time every $\sim 10^9$ yr. 

The limit on the cooling rate below a temperature $T_{\rm min}$
inferred from X-ray observations is $< 20 \%$ of the 
mass cooling rates cited in the past (Fabian 2002).
That is, X-ray emission from gas below $T_{\rm min}$ is absent
in present X-ray telescopes.  
This minimum temperature is different for different clusters, but
it is $\sim 1/3$ of the background temperature 
(Kaastra et al. 2003; Peterson et al.\ 2003b).
Despite the missing X-ray emission reported by these studies,
there are other indications that cooling to temperatures much below
$T_{\rm min}$ does occur, but on a much lower rate than that
inferred by taking the CF age to be equal to the cluster age.
Wise, McNamara, \& Murray (2003) and McNamara, Wise, \& Murray (2003),
for example, find the cooling rate within $r \sim 30 \kpc$ of the CF 
cluster A 1068 to be about equal to the star formation rate there
$\sim 20-70 M_\odot \yr^{-1}$.
Although the limits on the mass cooling rates cited above are within 
the values expected in the moderate CF model, a much better
agreement with observations can be achieved if some non-radiative 
cooling of the X-ray gas is added to the moderate CF model
(Soker, Blanton, \& Sarazin 2003, hereafter SBS).
The thermal energy released in cooling the hot gas is delivered
to the cold clouds, and is radiated as optical/UV line emission at
$\sim 10^4-10^5$ K.
This energy of the hot gas can be delivered to the cool gas by mixing 
(e.g., Oegerle et al.\ 2001; Fabian et al.\ 2001, 2002; 
Johnstone et al.\ 2002; Bayer-Kim et al.\ 200), or via heat conduction 
(SBS)\footnote{This local heat conduction model is 
completely different from large scale ($\gtrsim 100$ kpc) 
heat conduction models aiming at heating, rather than cooling, the
X-ray emitting gas in the center of CF clusters.}

In the scenario proposed by SBS the heat conduction is regulated by 
reconnection between the magnetic field lines in cold ($\sim 10^4$ K) 
clouds and the field lines in the ICM.
A narrow conduction front is formed, which, despite the relatively
low temperature, allows efficient heat conduction from the hot ICM
to the cold clouds.
The reconnection between the field lines in cold clouds and those
in the ICM occurs only when the magnetic field
in the ICM is strong enough.
This occurs only in the very inner regions of CF clusters,
at $r \sim 10-30$ kpc.
While in the scenario of SBS the non-radiative cooling starts with 
the reconnection of magnetic field lines from the pre-existing 
cold clouds to the X-ray emitting gas, in the present paper I 
examine heat conduction within pre-existing magnetic flux loops 
of X-ray emitting gas. 
The non-radiative cooling starts when small portion of the loop is 
radiatively cooling to $\sim 10^4 \K$. 

Magnetic flux loops in CF clusters were studied before, e.g.,
in relation to global heat conduction (Norman \& Meiksin 1996), in the
amplification of magnetic fields in the center of CF clusters
(Godon, Soker, \& White 1998), and the formation of multiphase
ICM (Zoabi, Soker, \& Regev 1996, 1998; hereafter ZSR). 
In a recent paper Kaastra et al.\ (2003) note the similarity of
the differential emission measure in the clusters they study
to that of coronal loops in the sun.  
The behavior of loops found by ZSR, the
finding of Kaastra et al.\ (2003), and the dynamically important
magnetic field in the inner regions of cluster CFs 
(Eilek \& Owen 2002; Taylor, Fabian, \& Allen 2002), motivate me
to study non-radiative cooling within magnetic flux loops.
This paper is not aiming at completely suppressing CF, as 
there are many evidence for cooling to low temperatures, e.g.,
Kaastra et al.\ (2003) find radiation from gas at temperatures of 
$\lesssim 10^6  \K$ in some clusters, Edge et al.\ (2002;
see also Edge \& Frayer 2003; Salome \& Combes 2003a,b;
Edwards 2003) find massive cold reservoir in the form of
molecular gas, probably from cooling ICM, and the star formation
rate cited above (Wise et al.\ 2003).
The main goal is to study another aspect, i.e., non radiative cooling
of X-ray emitting gas in magnetic flux loops,
and incorporate it into the moderate cluster CF model.
As such, this paper is complementary to SBS.

\section{The Rapid Radiative Cooling Phase of the Loop}

ZSR consider an elliptically shaped magnetic flux loop moving 
radially inward in a cluster CF. 
They follow only the two end points of the loop, assuming that
they evolve in series of hydrostatic equilibria, i.e., magnetic
tension $T_B$ balances buoyancy $F$ at each of the two end
points of the loop. 
This dictates that the density in the upper segment of the 
loop $\rho_u$, is lower than that of its surroundings, 
and the density of the lower segment of the loop $\rho_d$,
is higher than that of its surroundings. 
ZSR consider radiative cooling, but neglect heat conduction, 
even along the magnetic field lines. 
They base this assumption on a possible reduction in the heat 
conduction coefficient along magnetic fields lines
(Pistinner \& Eichler 1998). This reduction in heat conduction is
not efficient when the magnetic pressure becomes comparable to the
thermal pressure (Pistinner \& Eichler 1998). 
ZSR take the magnetic pressure to increase as density increases, 
and take a pressure equilibrium between the loop
and its surroundings.
As the loop evolves and falls toward the cluster center it cools 
with the lower-segment temperature $T_d$ falling faster
than the upper-segment temperature $T_u$. 
The loop is stretched in the radial direction and gets narrow
on its sides (see fig. 1). 
ZSR find that because the magnetic tension prevents the dense lower
segment of the loop from falling toward the cluster center, 
this lower segment reaches a phase of catastrophic cooling. 
For the parameters used by ZSR the catastrophic cooling occurs
when the lower segment is inside a cluster radius of $r<20 \kpc$,
but the upper regions of the loop extend to much larger distances, 
depending on its size.  

In the present paper I examine possible effects of heat 
conduction along magnetic field lines inside the loop. 
The two relevant time-scales are the conduction time scale along the
loop, from the upper hot segment to the cooling lower segment,
and the radiative cooling time at the denser lower segment.
Heat conduction proceeds on a time scale of
$\tau_{\rm con} \simeq ( 5 n {\rm k} T/2)(\chi T^{7/2} L^{-2})^{-1}$,
where the first parenthesis is the enthalpy per unit volume, and the
second parenthesis is the heat loss per unit volume per unit time
because of heat conduction.
The temperature $T$ is some average inside the loop, which at early
stages is close to both $T_u$ and $T_d$.
Before the lower segment of the flux loop rapidly cools to
$T_d \ll 10^7 \K$, the temperature gradient inside the loop
is shallower than $T_u/L$, and the heat conduction time is longer 
than the one given above. 
Here, $\chi$ is (nearly) a constant defined such that $\chi T^{5/2}$
is the heat conduction coefficient, $n$ is the total number density,
and $L$ the width of the conduction front, taken here to be the
length of the magnetic flux loop (see figure 1).
\begin{figure}
\centering\epsfig{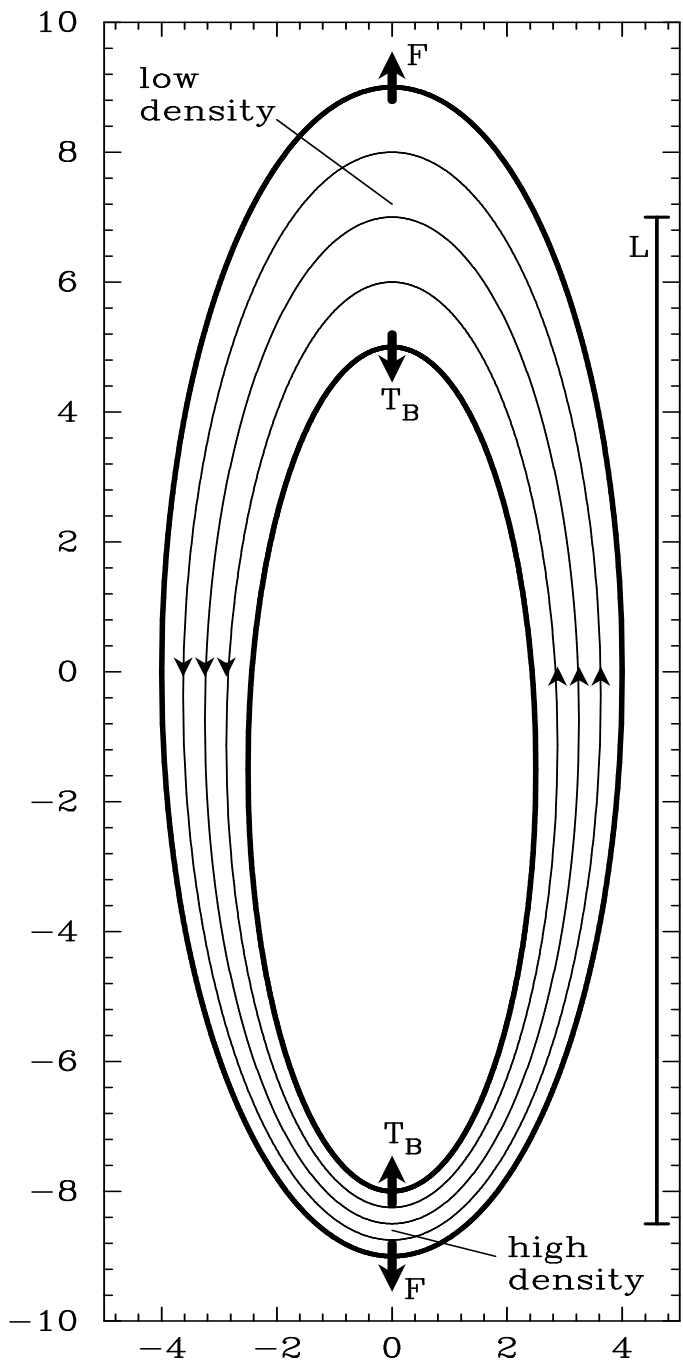}
\vskip 0.2 cm
\caption{A schematic structure of a magnetic flux loop.
Units on axes are in kpc.
Thick arrows depicted forces: $T_B$ magnetic tension and $F$
buoyancy. The thin ellipses are magnetic field lines, and $L$
is the length of the loop.
The lower segment is denser and cooler than its surroundings,
while the upper segment density is below that of its surroundings. 
After the lower segment cools to very low temperatures, efficient
heat conduction proceeds from the upper segment to the lower one 
along magnetic field lines, hence reducing the luminosity 
from X-ray emitting gas.}
\end{figure}
Scaling with typical values gives  
\begin{equation}
\tau_{\rm con} \simeq 4 \times 10^9
\left( \frac {n_e} {0.04 \, {\rm cm}^{-3}} \right) 
\left( \frac {T} {10^7 \, {\rm K}} \right)^{-5/2} 
\left( \frac {L} {10 \, {\rm kpc}} \right)^{2} \, {\rm yr} \, , 
\end{equation}  
where $n_e$ is the electron density.
At early stages, the time $\tau_{\rm con}$ is the time required to 
transfer heat from the upper segment to the lower one to prevent
it from cooling to very low temperatures.

The isobaric radiative cooling time of the lower segment 
in the temperature range 
$2 \times 10^5 \lesssim T \lesssim 4 \times 10^7$ K is
(see cooling curve in, e.g., Gaetz, Edgar, \& Chevalier 1988)
\begin{equation}
\tau_{\rm rad} \simeq 2 \times 10^8
\left( \frac {n_e} {0.04 \, {\rm cm}^{-3}} \right)^{-1} 
\left( \frac {T_d} {10^7 \, {\rm K}} \right)^{3/2} \, {\rm yr} \, .
\label{eq:trad}
\end{equation}  
The heat conduction will not prevent the lower segment from
cooling to lower temperatures when the condition 
$\tau_{\rm con} > \tau_{\rm rad}$ is met. 
Using equations (1) and (2), and substituting $T_d$ for $T$
(this is reasonable before the catastrophic cooling of the lower 
segment starts), the condition for rapid cooling of the lower segment reads 
\begin{equation}
T_d \lesssim 1.7 \times 10^7 
\left( \frac {T n_e} {4 \times 10^5 \K \cm^{-3}} \right)^{1/3} 
\left( \frac {L} {10 \kpc} \right)^{1/3} \K ,
\end{equation}  
where I took a constant pressure (expressed as $T n_e$, and scaled with
$P=10^{-10} \erg \cm^{-3}$), such that the condition is on the temperature 
of the lower loop's segment. 
It is evidence that the temperature when rapid cooling starts depends 
weakly on the external pressure and on the loop size. For the typical
parameters used here, it is in the range of the minimum
temperature of X-ray emitting gas deduced from observations
(Peterson et al.\ 2003a; Kaastra et al.\ 2003).

To summarize this section, there are two conditions for the 
catastrophic cooling of the lower segment of a flux loop to occur.
First, the magnetic field must be strong, close to equipartition,
in order to prevent the dense segment from falling in too fast.
This condition is also true for the scenario of SBS, where the
strong magnetic field is required to initiate reconnection between
cold clouds and the ICM. 
Second, the lower segment temperature must reach the value given
by equation (3).
Both conditions are likely to be met only near the cluster's center. 

\section{The Quasi-Static Heat Conduction Front Phase}

Borkowski, Balbus, \& Fristrom (1990) numerically study the 
evolution of a conduction front between a gas at $T_u \sim 10^6 \K$
and cooler gas at $10^4 \K$. 
They start with a sharp boundary between the two media. 
After a short period of time, during which the conduction front
widens, a quasi-static front is developed. 
In the third phase$-$the condensation phase$-$the entire front 
radiatively cools down. 
During the quasi-static front phase Borkowski et al.\ (1990) find
the mean temperature of different ions to be
$T_{\rm ions} \simeq 1-4 \times 10^{5} \K$,
i.e., much below the temperature of their hot phase which is
$T_u \sim 5-10 \times 10^5 K$.
I consider a similar effect here, where the mean temperature
is low, with most radiation emitted at much lower
temperatures than $T_u$.
The initial situation here is not a sharp boundary between the
hot and cold gases, but rather the lower segment of the loop
cools from a high temperature.
Therefore, I expect the quasi-static front phase to be
established quite early.

An accurate solution for the temperature distribution along
a static magnetic flux tube was derived analytically by
Rosner, Tucker, \& Vaiana (1978) and numerically by 
Aschwanden \& Schrijver (2002).
Both works consider heating as well as heat conduction and
radiative cooling.
Kaastra et al.\ (2003) when comparing their spectra to
the results of Aschwanden \& Schrijver (2002) assume the existence
of a heating source, but they don't specify its nature.
I here neglect heating, because I assume sporadic heating events
in the cluster, which most loops do not encounter during
their catastrophic cooling.
The results of Rosner et al.\ (1978) can be reasonably fitted
along most of the tube by $T \propto s^n$ with $n=1/3$, where
$s$ is the coordinate along the flux tube, i.e., along the
magnetic field lines, and $s=0$ is the location of the already
cold gas; here it is the lower segment of the loop.
Only at the highest temperature segment the temperature pofile
becomes shallower.
Practically the temperature of the lower segment
after its catastrophic cooling can be set to $T=0$,
although it is $\sim 10^4 \K$.
The numerical results of Aschwanden \& Schrijver (2002) depend
on the assumptions used, but are crudely fit by
$n \simeq 0.3$.
These results fit the quasi-static heat conduction
phase discussed here, as I now show.

A simple balance between radiative cooling and heat conduction at
each point along the flux loop reads (in the temperature range 
$2 \times 10^5 \lesssim T \lesssim 4 \times 10^7$) 
\begin{equation}
\frac{d}{ds} \left( \chi T^{5/2} \frac{dT}{ds} \right) =
n_e^2 \Lambda_e,
\end{equation}
where $\Lambda_e$ is the cooling function, with $\Lambda \propto T^{-1/2}$
in the relevant temperature range.
To obtain a simple analytical solution I take a constant pressure
along the flux tube, such that $n_e \propto T^{-1}$.
This allows the last equation to be written as
\begin{equation}
\frac{d^2 (T^{7/2})}{ds^2} = A_1 T^{-5/2},
\end{equation}
where$A_1$ is a constant composed of the different coefficients
in equation (4). 
This can be integrated once to yield
\begin{equation}
T^{5/2} \frac {d T}{ds} =2\left(\frac{A_1}{7}\right)^{1/2} (T + C_1)^{1/2},
\end{equation}
where $C_1$ is a constant of integration to be determined by
boundary conditions.
The left hand side of the last equation is the heat flux.
Rosner et al.\ (1978) take this flux to be equal to zero at
the higher temperature end of the loop.
However, we see that such a boundary condition here requires
$C_1 = -T_u$, which is not possible for lower temperatures.
A constant heating along the flux tube can make this solution possible,
by balancing the slow cooling of the high temperature (and low density)
segment. Otherwise heat conduction balances radiative cooling
and the heat flux can't vanish, and on the hottest end of the loop
no real steady state situation exists.

Equation (4) can analytically be integrated in principle.
However, the expression gives $s$ as function of $T$ and it is
too complicated for practical uses.
Instead I look for a solution in the form $T \propto s^n$.
I start by taking a minimum heat flux, which vanishes when
$T=0$. This gives $C_1=0$.
The solution reads (with $s=0$ where $T=0$)
\begin{equation}
T=A_2 s^{1/3}, 
\end{equation}
where $A_2=6^{1/3} (A_1/7)^{1/6}$.
As this is close to the solution obtained by Rosner et al. (1978),
I consider it a satisfactory approximation.
$C_1>0$ gives a lower value of $n$.
For the extreme case of $C_1 \gg T$ in the entire range
(which implies a huge heat flux, hence it is not appropriate
here where no heating is considered), the solution gives $n=2/7$.
The variation of $T$ with $s$ is somewhere between these two extremes,
i.e., $(2/7)< n < (1/3)$, but very likely close to $T \propto s^{1/3}$.
For a constant pressure along the loop, the relative emission
per unit volume is given by $\dot e = n_e^2 \Lambda_e \propto T^{-5/2}$.
For a temperature given by $T \propto s^n$, and a constant pressure
along the loop, the luminosity fraction of gas at temperature
$<T$ is given by (assuming constant loop's cross section)
\begin{equation}
\eta(T) = \left( \int_0^T T^{-5/2} ds \right)
\left( \int_0^{T_u} T^{-5/2} ds \right)^{-1} =
\left( \frac{T}{T_u} \right)^{\frac{1}{n}-\frac{5}{2}},  
\end{equation}
where in the second equality I substituted $ds =T^{(\frac{1}{n}-1)} dT/n$,
and preformed the integration.
Note that $T_u$ is not the maximum temperature in the CF cluster
$T_{\rm max}$, but rather the maximum temperature inside
the specific magnetic flux loop under consideration, and
$T_u < T_{\rm max}$. 

For $n=1/3$ the luminosity of gas at temperature $<T$ goes as
$\eta \propto T^{1/2}$.
Lower values of $n$ result in more emission from high temperature
gas, but the higher value of $n \simeq 1/3$ seems appropriate for
the following reason.
I assumed above that the pressure along the flux tube is constant.
However, in clusters the upper segment of the loop will be further
out in a lower pressure environment.
Outside $r \sim 25 \kpc$ in A~2052 (Blanton et al.\ 2001)
and $r \sim 10 \kpc$ in Hydra~A (David et al.\ 2001),
the pressure drops as $P \sim r^{-1}$.
This implies a pressure drop by a factor of $\sim 1.5$
between the two ends of a loop with $L\sim 10 \kpc$.
The lower pressure, hence density, at the upper loop segments
implies less emission from the hot gas compared
to the assumed constant pressure case.
Therefore, taking the higher value of $n$ in the range given above
is justified.
For the $n=1/3$ case, hence $\eta=(T/T_u)^{1/2}$,
$\sim 30 \%$ of the luminosity comes from gas at $T<0.1 T_u$,
and $\sim 50 \%$ from, gas at $T < 0.25 T_u$.
For example, if $T_u = 1 \kev$ (recall that $T_u$ is not the
maximum temperature in the CF cluster), then $\sim 30 \%$
($\sim 50 \%$) of the luminosity comes from gas at
temperatures $T<0.1 \kev$ ($T< 0.25 \kev$).
Overall, heat conduction within a magnetic flux loop
can reduce the fraction of energy radiated in the
X-ray band by a factor of $\sim 1.5-2$.

\section{Discussion and Summary}

I found above that heat conduction within a magnetic flux loop
can reduce the fraction of energy radiated in the
X-ray band by a factor of $\sim 1.5-2$, i.e.,
the X-ray luminosity is $\sim 50- 70 \%$ of the value expected
without heat conduction.
Cooling of X-ray emitting gas by conduction to cold gas
was proposed to occur via mixing (e.g., Oegerle et al.\ 2001;
Fabian et al.\ 2001, 2002; Johnstone et al.\ 2002; 
Bayer-Kim et al.\ 200) and by local heat conduction regulated 
by magnetic field reconnection (SBS).
In the later two processes the temperature gradient is much higher
than the one inside the loop studied here, and hence these
mechanism are more efficient in reducing X-ray emission.
However, the question is whether these processes occur, and how often.
The magnetic flux loop model studied here is motivated
by previous suggestions that many such loops exist in
the ICM of CF clusters and that they play an active
role in cooling and/or dynamics of the gas
(e.g., Zoabi et al.\ 1996; ZSB; Kaastra et al.\ 2003).

SBS estimate that in their heat conduction scenario via magnetic
field reconnection the total line emission expected for gas
cooling from $\sim 10^7$ K to $\sim 10^6$ K might be factor of
$\sim 30-100$ lower than that expected in older versions of the CF
model, where only line emission is considered for the cooling
process, and where the cooling rates are much higher.
This factor comes from a $\sim 10$ times lower mass cooling rate of 
the hot gas in the moderate CF model (Soker et al.\ 2001),
and from another factor of $\sim 3-10$ for cooling via heat conduction.
This much lower expected radiative cooling rate at low
X-ray emitting temperatures is compatible with limits
inferred from observations.
The observational limit, which is $\lesssim 10-20 \%$ of the expected
cooling rate in old versions of the CF model (Fabian 2002;
Peterson et al.\ 2003a), can be met if the reduction due
to heat conduction is only a factor of $\sim 2$ as found here,
with another factor of $\sim 10$ from the moderate CF model.
Possibly, the more significant role of magnetic flux loops
is to stir the ICM in the center of CF clusters, by reconnection
and motion caused by magnetic tension (Zoabi et al.\ 1996; ZSR).
This stirring, where some portions of loops falls fast
and others buoyant outward, can lead to more reconnection
between magnetic flux loops (Zoabi et al.\ 1996), enhancing
the efficiency of non-radiative cooling via local heat conduction
as discussed by SBS.
                
If the general scenario discussed here and/or in SBS holds,
then a CF model with the following ingredients emerges.
\newline
{\bf (1) CF does occur.}
The evidences for CF in many clusters were summarized by
Soker et al.\ (2001) prior to the new results of
the {\it Chandra} and XMM-Newton X-ray telescopes.
The new results show that the cooling rate to low temperatures
is much lower than the expected values in the old versions of
the CF model, but they do show that some cooling occurs
(e.g., a few clusters in Kaastra et al.\ 2003).
This is compatible with the moderate CF model.
On the other hand, in most CF clusters the new X-ray observations
didn't reveal any violent heating events, e.g., major mergers
or strong shocks, and the ICM structure is not consistent
with global heat conduction from the outer hot massive clusters
regions.
The X-ray observations do show heating, e.g., bubbles and
sound waves (e.g., Fabian et al.\ 2003), but these can't compensate
for the cooling in many cases (e.g., Kawano et al.\ 2003).
Observations using the high spatial resolution of {\it Chandra}
show most activity to occur in the very inner region, on a scale
much smaller than the cooling radius (where radiative cooling
equals the cluster age), compatible with much lower mass cooling
rate.
All these suggest that the new X-ray observations actually
strengthen the CF model, although a moderate one, rather
than ``kill'' it.
\newline
{\bf (2) Non-steady state evolution.}
Soker \& David (2003) show how a non-steady state situation
can account for the new observations. 
Such a non-steady state evolution is expected in  
intermittent-heating models, with long enough time intervals.
The gas is heated up, and then cools. 
Most of the heated gas does not cool to low temperatures before 
the next major heating event. Soker \& David claim that
no fine tuning is required. 
More accurate analysis may show that this non-steady state evolution
can hold for heating events with shorter intervals as well.
\newline
{\bf (3) Non-radiative cooling of X-ray emitting gas.}
Although not strictly a necessary ingredient, non-radiative cooling 
of the X-ray emitting gas after it cooled already by a factor 
of $\sim 2-5$ from its initial temperature, makes the agreement of 
the moderate CF with observations much better (SBS; this paper). 
The initial radiative cooling is required, most likely, to enhance
by a factor of $\gtrsim 10$ the magnetic pressure relative to the
thermal pressure (SBS).
The magnetic fields play a crucial role in the
scenario discussed here, and in the more efficient non-radiative
cooling scenario studied by SBS.
\newline
{\bf (4) A corona-like ICM.}
The outflow (buoyant bubbles and magnetic flux loops) and inflow,
the strong magnetic fields, the possible importance of local heat
conduction, radiative cooling in optically thin plasma, the presence
of X-ray emitting gas together with optically emitting gas, all
these and more, make the ICM in CF clusters similar in some aspects
to the gas above the solar (and similar stars) photosphere,
i.e., photosphere to corona regions.
The similarity with respect to some processes relevant to
magnetic field reconnection and optical line emission
was mentioned by Jafelice \& Friaca (1996) and Godon
et al.\ (1998).
Sparks \& Donahue (2003) note the morphological similarity between
filaments in M 87 and the solar atmosphere. 
Godon et al.\ (1998) discuss magnetic field amplification in
CF clusters and the possibility of ``cluster flares'', and
use hints from solar flares to project on the expected
nature of violent magnetic reconnection in CF clusters.
Some aspects of magnetic flux loop dynamics where studied
by Zoabi et al.\ (1996) and ZSR.
Kaastra et al.\ (2003) extend the comparison of ICM in CF cluster
with stellar coronae to explain the X-ray spectra they
observed in CF clusters.


\acknowledgements
This research  was partially supported by a grant from the
Israel Science Foundation.


\begin{references}
 
\reference{} Aschwanden, M. J., \& Schrijver, C. J. 2002, ApJS, 142, 269
 
\reference{}  Bayer-Kim C. M., Crawford C. S., Allen S. W.,
  Edge A. C., Fabian, A. C. 2002, MNRAS, 337, 938
 
\reference{} Binney, J.
    2003, in ``The Riddle of Cooling Flows in Galaxies and Clusters 
     of Galaxies'', eds. Reiprich, T. H., Kempner, J. C., \& Soker, N., 
      Charlottesville, VA, USA, May 31-June 4, 2003 (astro-ph/0310222) 

\reference{} Binney, J., \& Tabor, G. 1995, MNRAS, 276, 663

\reference{} Blanton, E. L., Sarazin, C. L., \& McNamara, B. R.
2003, ApJ, in press (astro-ph/0211027) 

\reference{} Blanton, E. L., Sarazin, C. L., McNamara, B. R., \&
Wise, M. W. 2001, ApJ, 558, L15 
 
\reference{} Borkowski, K. J.,  Balbus, S. A., \& Fristrom, C. C. 1990,
    ApJ, 355, 501
 
\reference{} Br\"uggen, M., \& Kaiser, C. R. 2001, Nature, 418, 301

\reference{} Ciotti, L., \& Ostriker J. P. 2001, ApJ, 551, 131
 
\reference{} David, L. P., Nulsen, P. E. J., McNamara, B. R.,
 Forman, W., Jones, C., Ponman, T., Robertson, B., \& Wise, M.
 2001, ApJ, 557, 546
 
\reference{} Edge A. C., \& Frayer, D.T. 2003, ApJ, 594, L13
  
\reference{} Edge, A. C., Wilman, R. J., Johnstone, R. M.,
   Crawford, C. S., Fabian, A. C., \& Allen, S. W. 2002, MNRAS, 337, 49

\reference{} Edwards, L. O. V.
    2003, in ``The Riddle of Cooling Flows in Galaxies and Clusters 
     of Galaxies'', eds. Reiprich, T. H., Kempner, J. C., \& Soker, N., 
      Charlottesville, VA, USA, May 31-June 4, 2003  

\reference{} Eilek, J. A., \& Owen, F. N. 2002, ApJ, 567, 202
 
\reference{} Fabian, A. C. 1994, ARA\&A, 32, 277 

\reference{} Fabian, A. C. 2003, MNRAS, 344, L27
 
\reference{} Fabian, A. C. 2002,  in Galaxy Evolution: Theory and
Observations, eds. V. Avila-Reese, C. Firmani, C. Frenk, \& C. Allen,
RevMexAA SC, in press (astro-ph/ 0210150)

\reference{} Fabian, A. C., Allen, S. W., Crawford, C. S.,
    Johnstone, R. M., Morris, R. G., Sanders, J. S., \& Schmidt, R. W.
     2002, MNRAS, 332, L50   

 
\reference{} Fabian, A. C.,  Mushotzky, R. F.,  Nulsen, P. E. J.,
   \&  Peterson, J. R. 2001, MNRAS, 321, L20 
 
\reference{} Fabian, A. C., Sanders, J. S., Allen, S. W., 
   Crawford, C. S., Iwasawa, K., Johnstone, R. M., Schmidt, R. W.,
   \& Taylor, G. B. 2003, MNRAS, 344, L43
 
\reference{} Gaetz, T. J., Edgar, R. J., \& Chevalier, R. A.
            1988, ApJ, 329, 927
 
\reference{} Godon, P., Soker, N., \& White, R. E., III 1998,
    AJ, 116, 37 

\reference{}  Heinz, S.; En{\ss}lin, T.
     2003, in ``The Riddle of Cooling Flows in Galaxies and Clusters 
     of Galaxies'', eds. Reiprich, T. H., Kempner, J. C., \& Soker, N., 
      Charlottesville, VA, USA, May 31-June 4, 2003  


\reference{} Jafelice, L. C., \& Friaca, A. C. S. 1996, MNRAS, 280, 438
 
\reference{} Johnstone, R. M., Allen, S. W., Fabian, A. C., \&
  Sanders, J. S. 2002, MNRAS, 336, 299
  
\reference{} Kaastra, J. S. et al.\ 2003, A\&A, in press (astro-ph/ 0309763).
 
\reference{} Kahn, S. M., Peterson, J. R., Paerels, F. B. S., Xu, H.,
    Kaastra, J. S., Ferrigno, C., Tamura, T., Bleeker, J. A. M., 
    \& Jernigan, J. G. 2003,   (astro-ph/0210665)

\reference{} Kawano, N., Ohto, A., \& Fukazawa, Y. 2003, PASJ, 55, 585 

\reference{} Mathews, W. G., Brighenti, F., Buote, D. A., \& Lewis, A. D. 
           2003, ApJ, 596, 159
  
\reference{} Mazzotta, P., Edge, A. C., \& Markevitch, M. 2003, ApJ, 596, 190
  
\reference{} McNamara, B. R., Wise, M. W., \& Murray, S. S. 2003,
 ApJ, in press (astro-ph/0310035) 
 
\reference{} Norman, C., \& Meiksin, A. 1996, ApJ, 468, 97 
 
\reference{} Nulsen, P.E.J., David, L., McNamara, B., Jones, C., Forman, W. ,
	   Wise, M. 2002, ApJ, 568, 163.
 
\reference{} Oegerle, W. R., Cowie, L., Davidsen, A., Hu, E.,
   Hutchings, J., Murphy, E., Sembach, K., \& Woodgate, B. 
   2001, ApJ, 560, 187 
 
\reference{} O'Sullivan, E., \& Vrtilek, J. M. 
     2003, in ``The Riddle of Cooling Flows in Galaxies and Clusters 
     of Galaxies'', eds. Reiprich, T. H., Kempner, J. C., \& Soker, N., 
      Charlottesville, VA, USA, May 31-June 4, 2003 (astro-ph/0310099) 
    
\reference{} Peterson, J. R., Kahn, S. M., Paerels, F. B. S.,
    Kaastra, J. S., Tamura, T., Bleeker, J. A. M., Ferrigno, C.,
     \& Jernigan, J. G. 2003a, ApJ, 590, 207 

\reference{} Peterson, J. R., Kahn, S. M., Paerels, F. B. S.,
    Kaastra, J. S., Tamura, T., Bleeker, J. A. M., Ferrigno, C.,
     \& Jernigan, J. G. 2003b, in ``The Riddle of Cooling Flows 
     in Galaxies and Clusters of Galaxies'', 
       eds. Reiprich, T. H., Kempner, J. C., \& Soker, N., 
      Charlottesville, VA, USA, May 31-June 4, 2003, 
 
\reference{} Pistinner, S. L., \& Eichler, D. 1998, MNRAS, 301, 49
 
\reference{} Quilis, V., Bower, R. G., \& Balogh, M. L.
    2001, MNRAS, 328, 1091
 
\reference{} Rosner, R., Tucker, W. H., \& Vaiana, G. S. 1978, ApJ, 220, 643
 
\reference{} Ruszkowski, M., \& Begelman, M. C. 2002, ApJ, 581, 223
   (astro-ph/0207471) 

\reference{} Salome, P., \& Combes, F. 2003a
  in ``The Riddle of Cooling Flows in Galaxies and Clusters of Galaxies'', 
  eds. Reiprich, T. H., Kempner, J. C., \& Soker, N., 
  Charlottesville, VA, USA, May 31-June 4, 2003  
 
\reference{} Salome, P., \& Combes, F. 2003b, A\&A in press (astro-ph/0309304)
 
\reference{} Smith, D. A., Wilson, A. S., Arnaud, K. A., Terashima, Y.,
   \& Young, A. J. 2002, ApJ, 565, 195 

\reference{} Soker, N., Blanton, E. L., \& Sarazin, C. L. 2003,
  in ``The Riddle of Cooling Flows in Galaxies and Clusters of Galaxies'', 
  eds. Reiprich, T. H., Kempner, J. C., \& Soker, N., 
  Charlottesville, VA, USA, May 31-June 4, 2003 (astro-ph/0309633) (SBS) 

\reference{} Soker, N. \& David, L. P. 2003, ApJ, 589, 770

\reference{} Soker, N., White, R. E. III, David, L. P., \& McNamara, B. R.
    2001, ApJ, 549, 832 

\reference{} Sparks, W. B., \& Donahue, M. 2003  
  in ``The Riddle of Cooling Flows in Galaxies and Clusters of Galaxies'', 
  eds. Reiprich, T. H., Kempner, J. C., \& Soker, N., 
  Charlottesville, VA, USA, May 31-June 4, 2003 

\reference{} Taylor, G. B., Fabian, A. C., \&  Allen, S. W. 2002,
   MNRAS, 334, 769
 
\reference{} Tucker, W., \& David, L. P. 1997, ApJ, 484, 602

\reference{} Wilson, A. S., Young, A. J., \& Smith, D. A. 2003,
    in Active Galactic Nuclei: from Central Engine to Hot Galaxies,
    ASP Conf. Ser., eds. S. Colin, F. Combes, and I. Shlosman
     in press (astro-ph/0211541) 

\reference{} Wise, M. w., McNamara, B. R., \& Murray, S. S. 2003,
 ApJ, in press (astro-ph/0310033) 

\reference{}  Zoabi, E., Soker, N., \& Regev, O. 1996, ApJ, 460, 244 

\reference{}  Zoabi, E., Soker, N., \& Regev, O. 1998, MNRAS, 296, 579 (ZSR)

\end{references}
\end{document}